\documentclass[twocolumn,prb,aps,longbibliography,nobalancelastpage]{revtex4-2}
\usepackage{times}

\usepackage{amsmath,amsfonts,amssymb,graphicx,hyperref,epstopdf,xcolor,tikz,scalerel,natbib,array,gensymb}
\usepackage{mathptmx,textcomp,braket}
\usepackage[T1]{fontenc}
\usepackage[utf8]{inputenc}
\usepackage{multirow}
\usepackage{array}
\DeclareMathAlphabet{\mathpcal}{OMS}{zplm}{m}{n}

\definecolor{lime}{HTML}{A6CE39}
\DeclareRobustCommand{\orcidicon}
{
	\begin{tikzpicture} 
	\draw[lime, fill=lime] (0,0) circle [radius=0.15] node[white] {{\fontfamily{qag}\selectfont \tiny ID}};
	\draw[white, fill=white] (-0.0625,0.095) 	circle [radius=0.007];
	\end{tikzpicture}
	\hspace{-2.2mm}
}
\newcommand\orcidID[1]{\href{https://orcid.org/#1}{\orcidicon}}

\newcommand{\be}{\begin {equation}}
\newcommand{\ee}{\end {equation}}
\newcommand{\beqa}{\begin {eqnarray}}
\newcommand{\eeqa}{\end {eqnarray}}

\hypersetup{colorlinks,citecolor=blue,filecolor=black,linkcolor=blue,urlcolor=blue}
 
\begin{document}

\title{Probing topological phase transitions in Aubry-Andre-Harper model via high-harmonic generation}

\author{Nivash R\orcidID{0009-0004-3076-9192}}
\email[E-mail: ]{nivash1807@gmail.com}
\author{Jayendra N. Bandyopadhyay\orcidID{0000-0002-0825-9370}}
\author{Amol R. Holkundkar\orcidID{0000-0003-3889-0910}}

\affiliation{Department of Physics, Birla Institute of Technology and Science - Pilani, Rajasthan,
333031, India.}

%\date{\today}

\begin{abstract}
  
We study the high-harmonic generation in the Aubry-Andre-Harper (AAH) model. The modulating phase of the AAH model is used as a control parameter while preserving the chiral symmetry hosting the zero-energy edge states. The harmonic yield in a particular energy range exhibits a strong dependence on the control parameter with clear separation of the region of topologically trivial and nontrivial phases of the system. The threshold for the harmonic yield is found to serve as an all-optical tool for detecting topological phases. We extended our study with broken chiral symmetry by including the onsite potential. The introduction of the onsite potential lifts the degeneracy in the edge states, which affects the harmonic enhancement. Furthermore, it is also observed that the system's onsite strength can control the HHG yield.

\end{abstract}

\maketitle

\section{Introduction}
 
 The high-harmonic generation (HHG) has been extensively studied in the gas phase over the last decades; it gives us a promising route to produce highly tunable extreme-ultraviolet (XUV) pulses in the attosecond regime \cite{RevModPhys},  along with the opportunity to investigate the electron dynamics in atoms and molecules on their natural timescales \cite{smirnova2009,corkum2007}. With the advancement of technology, mid-infrared sources are opening new avenues in the strong field interaction with the solids, as we witnessed from the pioneering work of Ghimire et al. \cite{Ghimire2011}. The HHG from solids promises a compact source of XUV radiation and attosecond spectroscopy \cite{luu2015, Kruchinin_2018, Nicoals_21}. The HHG from the solids enabled us to explore the countless possibilities in condensed matter physics, like exploring the large-band-gap dielectrics \cite{luu2015}, retrieving the bandstructure \cite{Vampa_15}, the effect of vacancy defects \cite{Mrudul_20}, the petahertz current in solids \cite{luu2015}, HHG in graphene \cite{Yoshikawa_17, Mrudul_21} and Transition metal dichalcogenides \cite{yoshikawa_2019}, Bloch oscillations in solids \cite{schubert2014, Nivash_23}, and many more.

The invention of the quantum hall effect \cite{Thouless_82} escalated the research directions toward exploring the topological phases of matter. The topological material exhibits the insulating property in the bulk but has a conducting edge state, unaffected by the perturbations and defects. It enabled the researchers to explore the strong field phenomenon in topological condensed matter physics, especially the high-harmonic generation.

The HHG from the topological materials or, in general, solids can be understood in terms of intraband and interband current contributions. The motion of electron within a band contributes to the intraband emission with the photon energy below the band gap. However, the electron transitions from one band to the other band constitute the emission (above the band gap energy) caused by the interband current. In the case of the topological materials, an edge state in the band gap enhances the interband transitions, strengthening the high harmonic emission. In the later part of this manuscript, we will see how the characteristics of the HHG emission by the topological material might work as a diagnostic tool to distinguish between the topological phases. 

The contribution of the edge state in the HHG processes is widely studied in the context of the one-dimensional (1D) Su-Schrieffer-Heeger (SSH) model  \cite{Bauer2018, Bauer2019, Bian_Tao2022}, the extended SSH model \cite{Bera_23}, the AAH model \cite{Bian2022}, and the Kiteav model \cite{Pattanayak2022, Baldelli2022}. Furthermore, the HHG spectrum is an all-optical probe used to extract topological information using the helicity \cite{Silva2019, Heide2022} or circular dichroism \cite{Alexis_20} of the emitted photons in the Haldane model. With the above theoretical studies, whether the HHG spectrum shows unique characteristic features that exhibit topology remains to be seen. However, recently, an ab initio simulations of HHG \cite{Neufeld_23} shed light on the universal behavior of topological signatures in the HHG spectra.

This paper uses a control parameter in the Aubry-Andre Harper (AAH) model for the HHG processes and studies the topological phase transitions. In the past, this model was used to study the dynamics of particles in quasiperiodic systems. This AAH model can be experimentally accessible using photonic crystals \cite{Lahini2009, Lahini2015} and an optical waveguide array \cite{Lahini2009, Lahini2013, Zilberberg2012}. The quasiperiodicity of the AAH model in the form of a cosine modulation has its periodicity, which determines whether modulation is incommensurate (irrational) or commensurate (rational) with lattice spacing. Moreover, quasiperiodicity may be included in the terms of on-site and hopping. The incommensurate hopping modulation leads to Anderson-like localization \cite{Chong2015}, and the commensurate hopping modulation term leads to the appearance of zero-energy edge states \cite{Ganeshan2013, Cestari2016}. Interplay between both quasiperiodic modulations was studied in \cite{Chong2015}. The one-dimensional quasicrystals (QC) were found to be topologically nontrivial, with the periodically modulating parameter providing an additional degree of freedom, which can be mapped into the two-dimensional Hofstadter model, investigated theoretically \cite{Shuchen_2012, Zilberberg2012} and experimentally \cite{Lahini2009, Lahini2015}.   

The discrete symmetries of the system classify the properties of a topological insulator \cite{Ryu_2010,Chiu_16}. One of them is the chiral symmetry or sublattice symmetry (in the case of 1D TIs), which is responsible for the conservation of zero-energy edge states that are reflected in the anticommutation property between the chiral operator and the bulk Hamiltonian. The role of breaking of chiral symmetry in topological insulators has been investigated using HHG mechanism \cite{Bian_Tao2022}, and the emergence of opposite chirality in weyl semimetals has been studied recently under strong laser excitation \cite{Lv2021,Bharti_23}. Chirality has an attraction towards strong field dynamics. Recent studies claim that HHG can be a sensitive probe for chirality in higher-dimensional systems like the tellurium crystals \cite{chen_20}, the surface of silicon dioxide, and magnesium oxide \cite{Heinrich2021} by driving circularly polarised light. In this work, we explore the HHG by the AAH model with off-diagonal hopping while preserving the chiral symmetry. The HHG spectrum enhancement in the below band-gap reflects the radiation from the edge states in the trivial phase, and the absence of edge states leads to lower harmonic emission in the trivial phase. We observed that the interference of inter- and intra-band currents has a measurable impact on suppressing the total harmonic emission in the minimum band-gap region. The role of breaking chiral symmetry in the HHG spectrum has also been investigated. The distinction between the topological phases is commented upon by observing the harmonic yield with a change in the control parameter of the AAH model. This paper is organized as follows: the theoretical aspects are presented in Sec. \ref{sec2}, followed by the results and discussion in Sec. \ref{sec3}, and finally the summary in Sec. \ref{sec4}.

\section{Methodology}
\label{sec2}

\subsection{Model}

\begin{figure}[t]
\centering\includegraphics[width=1.0\columnwidth]{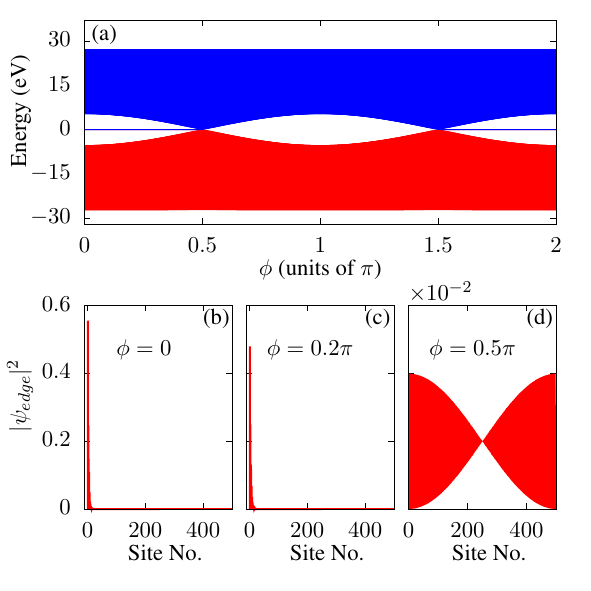}
\caption{The energy spectrum with 500 lattice sites as a function of $\phi$ (a). The red and blue colours represent the `Valence Band' (occupied states) and `Conduction Band' (unoccupied states), respectively. The lower plots show the probability distribution of the edge state in the AAH with off-diagonal hopping for three different values: $\phi$ = 0$\pi$ (b), $\phi$ = 0.2$\pi$ (c), and $\phi$ = 0.5$\pi$ (d).}
\label{fig1}
\end{figure}

We study here the 1D AAH with off-diagonal hopping \cite{Ganeshan2013} which is described by the following Hamiltonian  of the form,
\be
H_0 = \sum_{n=1}^{N} \eta[c_{n+1}^\dag c_{n} + h.c.]
\ee
where $c^\dag_{n}$ and $c_{n}$ are creation and anhilation operators, respectively. Here, $\eta$ is the hopping potential and has the form $\eta=\eta_0[1 + \lambda \cos(2\pi bn+\phi)]$, with $\lambda$ being the amplitude of quasiperiodicity and modulating phase factor $\phi$. The quasiperiodicity can be encoded through the cosine modulation with periodicity $1/b$, where $b$, the modulating frequency, controls the quasiperiodicity between commensurate and incommensurate cases. Depending on the value of $b$ (here in this work, $b$ takes the form $b=p/q=1/2$ with 'q' bands. The modulating parameter $\phi \in [0, 2\pi]$ appears as the additional degree of freedom (d.o.f) \cite{Lahini2012}, which affects the topological nature of the AAH model \cite{Lahini2015}. It allows us to map the 2D Hofstadter model, which describes electrons hopping on a 2D square lattice in a perpendicular magnetic field with $2\pi b$ magnetic flux quanta per unit cell \cite{Zilberberg2012}. In this work, we have set $t_0=0.5$ a.u., $\lambda=0.2$a.u. and $N=500$ lattice sites. The energy eigenvalues are plotted in Fig.\ref{fig1}(a). with respect to $\phi$, showing degenerate pairs of edge states in the regions: $0<\phi<0.5\pi$ and $1.5\pi<\phi< 2\pi$. Further, we plot the specific values of $\phi=0,0.2\pi$ and $0.5\pi$ by calculating the probability of zero energy eigenstate, which confirms the localization of edge states as shown in Fig. \ref{fig1}(c-e). The states are localised at the edge for $\phi=0$ in the probability plot, and localization strength decreases at 0.2$\pi$. With respect to $\phi$, localization strength decreases, and the state becomes delocalized at 0.5 $\pi$ and goes into the bulk part of the chain. The importance of localization and the presence of edge states will be discussed in detail in the upcoming section. 

This model preserves chiral symmetry \cite{Asboth_2016}, i.e., $ C^\dag H_0(k) C = - H_0(k)$, where $C=\sigma_z$ is the chiral opertor. The periodicity in the modulating frequency $b$ causes the model to dimerize \cite{Cao_2018}, i.e. there will be 'q' number of sublattices. Hence, the above equation is modified as: 
\be
H_0 = \sum_{n=1}^{N} [\eta_1 c_{n,A}^\dag c_{n,B} + \eta_2 c_{n,B}^\dag c_{n+1,A} + h.c.]
\ee
Here, there are two sublattices A and B with intracell hopping $\eta_1=\eta_0(1-\lambda \cos(\phi))$ and intercell hopping $\eta_2=\eta_0(1+\lambda \cos(\phi))$. This looks like a well-known SSH model and shares the same topology \cite{SSH}. The modulating phase $\phi$ alters the hopping strength, which causes the system to go from topologically nontrivial phase ($\eta_2 > \eta_1$) to topologically trivial phase ($\eta_2 < \eta_1$) phase, which can be distinguished by the topological invariant quantity, Winding number. 

\subsection{Coupling to an external field}

The energy eigenvalues and eigenstate are calculated by using the field-free Hamiltonian $H_0$. The laser pulse is polarized along the linear chain and the Hamiltonian becomes time dependent $H(t) = H_0 + H_I(t)$. The length-gauge interaction Hamiltonian \cite{Bian2022} reads us

\be
 H_I(t) = \sum_{n=1}^N n E(t) a_0 c_n c_n^{\dag}
\ee

where $a_0=2$ a.u. is the lattice constant, $E(t)=-{dA(t)}/{dt}$ is the electric field of the laser pulse, and the vector potential is $A(t)=A_0 \sin^2(\pi t/T)\sin(\omega_0 t)$. In this work, the amplitude of vector potential is kept as $A_0=0.1$ a.u., and the frequency of the laser is taken to be $\omega=0.0152$ a.u. ($\lambda=3\mu m$) for five optical cycles ($T=5\tau$), where $\tau$ is one optical cycle of the fundamental frequency ($\omega_0$).

The time-dependent Schrodinger equation is solved independently for the $m$ initially occupied eigenstates $\psi_m(t=t_i)$ of the field-free Hamiltonian by the Crank-Nicolson method to get the time-dependent wavefunction $\psi_m(t)$. Total current is thus obtained as:
\be
J_{\rm tot}(t) = \sum_m \langle \psi_m(t)\vert J \vert \psi_m(t) \rangle
\ee
where $J$, current operator is defined as
\be
J = -iea_0\sum_n \eta [c_n^{\dag}c_n - c_{n+1}^{\dag}c_n]
\ee
In order to extract the inter and intraband dynamics the time dependent wavefunction $\psi_m(t)$ is projected to each field free eigenstate \cite{Liu_17}.
\be
\vert \psi(t)\rangle = \sum_b \sum_m \alpha^b_m(t) \vert \phi^b_m \rangle
\ee
where b is the band index and m is the state index for every band. The intra and interband current can be written as
\be
J^{m}_{\rm intra}(t) = \sum_b \sum_{m,m^\prime} \alpha_m^{b^\ast} \alpha_m^{b} \langle \phi_m^b\vert J \vert \phi_m^b \rangle
\ee
\be
J^{m}_{\rm inter}(t) = \sum_{b,b^\prime} \sum_{i,j} \alpha_i^{b^\ast} \alpha_i^{b^\prime} \langle \phi_i^b\vert J \vert \phi_i^{b^\prime} \rangle
\ee
The total intra and interband current for each eigenstate m is calculated as
\be
J_{\rm tot}(t) = \sum_m J^{m}_{inter}(t) + J^{m}_{intra}(t)
\ee
The harmonic spectra are obtained by taking the Fourier transform harmonics of the time derivative current, which are given as:
\be S_\text{\rm tot}(\omega) = \Big|\mathpcal{F}_\omega[dJ_\text{total}/dt] \Big|^2 \ee 
where, $\mathpcal{F}_\omega[g(t)] = \int g(t) \exp[-i\omega t] dt$ is the Fourier transform of the time dependent function $g(t)$.
%The harmonic spectra are obtained by taking the fourier transform of $dJ_{tot}(t)/dt$.
%\be S_\text{total}(\omega) = \Big| \int j_\text{total}(t) e^{-i\omega t} dt \Big|^2 \ee
The total spectra [$S_{\rm tot}(\omega)$] of the emitted harmonics contains the summation of interband, intraband and interference of both \cite{Wang_2018,Nivash_24}
%\be S_\text{total}(\omega) = S_\text{inter}(\omega) + S_\text{intra}(\omega) + S_\text{intfer}(\omega) \ee 
\be S_\text{total}(\omega) = S_\text{inter}(\omega) + S_\text{intra}(\omega) + S_\text{intfer}(\omega) \ee 
where, 
%\be S_\text{inter,intra}(\omega) = \Big| \int j_\text{inter,intra}(t) e^{-i\omega t} dt \Big|^2,\label{specinter}\ee
\be S_\text{inter,intra}(\omega) = \Big|\mathpcal{F}_\omega[j_\text{inter,intra}]\Big|^2,\label{specinter}\ee
and,
\be S_\text{intfer}(\omega) = \mathpcal{F}_\omega^\ast[j_\text{inter}] \mathpcal{F}_\omega[j_\text{intra}] +  \mathpcal{F}_\omega^\ast[j_\text{intra}] \mathpcal{F}_\omega[j_\text{inter}],\label{intfer}\ee
The harmonic yield $Y$ is calculated using the relation, $Y = T^{-1} \int_{\omega_1}^{\omega_2}  S_\text{total}(\omega) d\omega$.

%\be
%\begin{split} 
%S_\text{intference}(\omega) = & \int j^\ast_\text{inter}(t) e^{i\omega t} dt \int j_\text{intra}(t) e^{-i\omega t} dt \\ & 
%+ \int j^\ast_\text{intra}(t) e^{i\omega t} dt \int j_\text{inter}(t) e^{-i\omega t} dt.
%\end{split}
%\ee

\begin{figure}[t]
\centering\includegraphics[width=1.0\columnwidth]{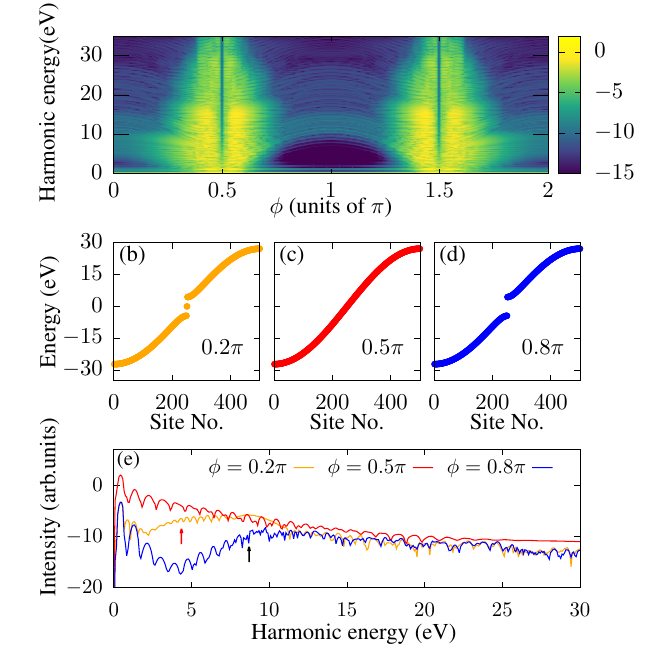}
\caption{(a) Shows that the $\phi$-dependent HHG spectrum corresponds to the energy spectrum in Fig. \ref{fig1}(a). The energy spectrum for the different topological phases. (b) $\phi=0.2\pi$ (topologically non-trivial), (c) $\phi=0.5\pi$ (gapless state), and (d) $\phi=0.8\pi$ (topologically trivial), along with the harmonic emission spectra, are presented in (e). The black arrow represents the minimum bandgap between the `Valence-Band' (VB) and `Conduction-Band' (CB) for $0.2\pi$ and $0.8\pi$, and the red arrow indicates the energy difference between the VB and edge states at $0.2\pi$.}
\label{fig2}
\end{figure}

\section{Results and Discussion}
\label{sec3} 

\subsection{Chiral preserving system}
The HHG spectrum calculated for all $\phi$ is shown in Fig. \ref{fig2}(a). We observed harmonic enhancements in the regions $\phi$ = (0.3-0.6)$\pi$ and $\phi$ = (1.4-1.7)$\pi$. In order to examine the harmonic enhancement, three specific $\phi$ values are choosen: (i) $\phi=0.2\pi$, Fig. \ref{fig2}(b) has a gapped state with the presence of edge state ($\eta_2 > \eta_1$); (ii) $\phi=0.5\pi$, gapless state ($\eta_2 = \eta_1$) in Fig. \ref{fig2}(c); and (iii) $\phi=0.8\pi$, Fig. \ref{fig2}(d) has a gapped state with no edge state ($\eta_2 < \eta_1$).

\begin{figure}[t]
\centering\includegraphics[width=0.9\columnwidth]{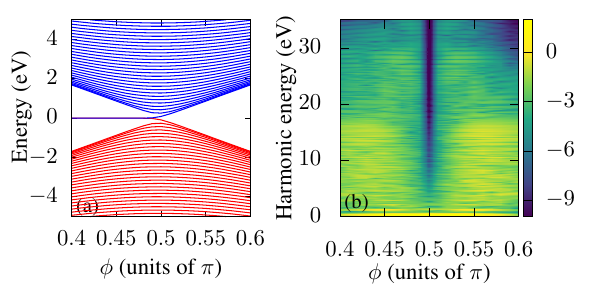}
\caption{An expanded view of the energy spectrum around topological phase transitions in region $\phi=0.4-0.6\pi$ is shown in (a), along with the respective harmonic spectrum (b).}
\label{fig3}
\end{figure}

Depending on the band gap (for each value of $\phi$), the harmonic spectra show unique behaviour. At $\phi=0.2\pi$ and $\phi=0.8\pi$, we can see a similar minimum band gap, where the minimum band gap is the energy difference between the last VB eigenstate and the first CB eigenstate. Intraband dynamics are dominant in the minimum band gap. The harmonic spectrum has a dip at $\phi=0.8\pi$, which contributes to destructive interference among multiple valence electrons driven by the laser field \cite{Bauer2019}. However, at $\phi=0.2\pi$, the harmonic yield increases after 4.3 eV, as marked by the red arrow in Fig. \ref{fig2}(e), which implies the zero energy states facilitate the interband transitions. On the other hand, $\phi=0.8\pi$ is a topologically trivial case with no mid-energy states (edge states) and thus the contribution of interband transitions after the electron surpasses the minimum band gap. As a consequence, there is a huge difference in the harmonic enhancement. At $\phi=0.5\pi$, the energy spectrum is gapless, and the transition happens at the edges of the two bands, resulting in an efficient harmonic yield. The above variation in the harmonic emission allows us to distinguish between the topologically trivial and nontrivial phases in the harmonic spectrum, as was also observed in previous studies of the HHG in chains.

The harmonic intensity is enhanced or decreased at certain $\phi$ values; this can be understood by studying the energy bands. If the bandgap is large (small), it is less (more) probable for an electron to excite to a conduction band, resulting in a lower (more) harmonic yield observed in the region $\phi=0.6\pi-0.8\pi$ ($\phi=0.4\pi-0.6\pi$), where mid-energy is enabling the transition to give a higher harmonic yield in the region of $\phi=0-0.4\pi$ and the absence of mid-energy state leads to a lower harmonic yield in $\phi=0.6\pi-1.3\pi$. This behaviour confirms the importance of the presence of an edge state, i.e. the non-trivial behaviour of the system. In order to understand the nature of the HHG spectra around the topological phase transition region, we have presented the band structure and harmonic spectrum around the range $0.4\pi-0.6\pi$ in Fig. \ref{fig3}(a-b). The zero-energy degenerate state splits when $\phi>0.48\pi$ and enters the bulk (showing delocalization). A similar harmonic yield is observed when $\phi=0.4\pi$ to $0.49\pi$ and $\phi=0.51\pi$ to $0.60\pi$, where the band gap reduces, enhancing the probability of transitions. A sudden drop in the harmonic emission is seen at $\phi=0.5\pi$.  

\begin{figure}[t]
\centering\includegraphics[width=1.0\columnwidth]{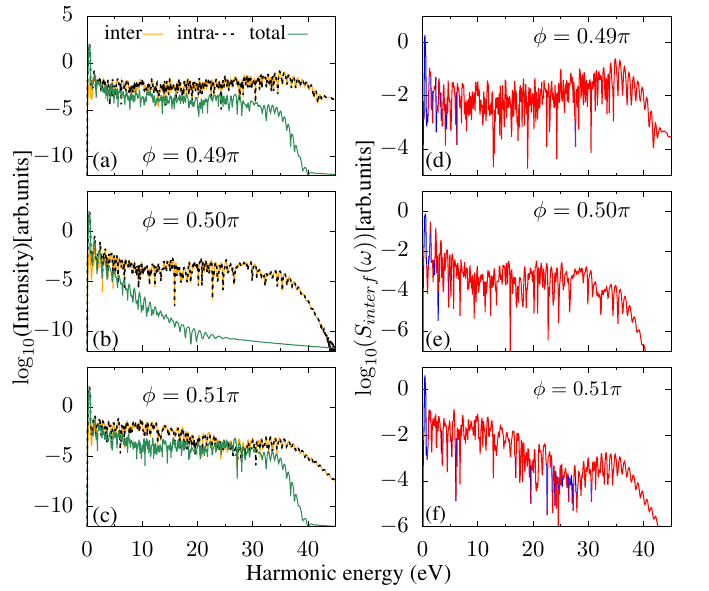}
\caption{HHG spectrum of total (black), intraband (green), and interband (red) are shown for $0.49\pi$ (a), $0.50\pi$ (b), and $0.51\pi$ (c). The corresponding interference term in the logarithmic scale are presented for $0.49\pi$ (d), $0.50\pi$ (e), and $0.51\pi$ (f).}
\label{fig4}
\end{figure} 

To further elucidate on this aspect, in Fig.\ref{fig4}(a-c), we have presented the HHG spectra with inter- and intraband harmonics when $\phi=0.49\pi-0.51\pi$. At $\phi=0.49\pi$, the intraband harmonics are comparable to the interband ones, and the interference effects have an impact after the cutoff, showing a cutoff around 35 eV. The interference term is plotted in logarithmic scale at Fig.\ref{fig4}(d-f). It has two contributions: enhancing (constructive interference) or suppressing (destructive interference) the total harmonic spectrum, and the negative (red) and positive (blue) values refer to destructive and constructive interference in the interference spectrum. The strong destructive interference happens at a cutoff at 0.49$\pi$ that correspondingly reduces the total harmonic spectrum. Since the total harmonic spectrum is a combination of interband, intraband, and interference of both can be obtained from Eq. (9). The interference term shows a huge impact on the total HHG spectrum at 0.5$\pi$, and it deviates from intra- and interband contributions that lead to strong destructive interference, as shown in Fig. \ref{fig4}(e). The total harmonic spectrum has a yield of around 10 eV at 0.51$\pi$, which correlates with the dominant destructive interference term in Fig. \ref{fig4}(e). In all the above cases, where the bandgap is small, the intraband and interband current oscillations occur simultaneously, where the interband current is raising and the intraband is falling, and vice versa. This helps us to conclude that both current oscillations are out of phase with each other in the time domain and have a similar intensity in the Fourier domain.

\subsection{System with broken Chiral symmetry}

\begin{figure}[t]
\centering\includegraphics[width=1.0\columnwidth]{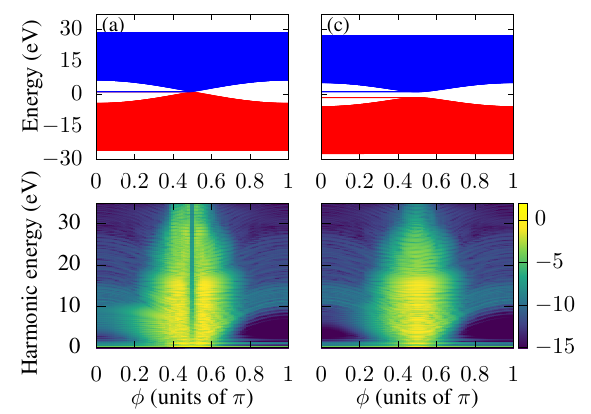}
\caption{Energy spectrum for including the onsite potential as a function of $\phi$. (a) $V_A=V_B=0.05$ a.u. (b) $V=0.05\cos(2\pi bn)$ a.u. (c,d) shows the harmonic spectrum vs. $\phi$, which corresponds to the energy spectrum in (a,b).}
\label{fig5}
\end{figure}

In this section, we study the impact of chiral symmetry breaking on the HHG spectrum. It can be broken either by introducing the onsite potential \cite{Bian_Tao2022} or next-nearest neighbour hopping potential \cite{chen_2014,Cao_2018}, which have been experimentally demonstrated in photonic crystals \cite{Jiao_2021}. Here, we specifically see the inclusion of the onsite potential terms as $H_ \mu = \sum_{n=1}^{N} [V_A c_{n,A}^\dag c_{n,A} + V_B c_{n,B}^\dag c_{n,B}] + H_0 $, where $V_A$ and $V_B$ are the strengths of the onsite potential, has an impact on this model as well as HHG. We first consider the case $V_A=V_B=0.05$. Fig.\ref{fig5}(c) shows the energy spectrum, which is similar to Fig.\ref{fig1}(a), but there is a lift in the degenerate edge states determined by the strength of the onsite strength and correspondingly a shift in the energy spectrum that is not going to change the physical aspects, and thus a similar harmonic spectrum is observed in Fig.\ref{fig5}(b) as in the off-diagonal hopping case. The energy and HHG spectra have a mirror symmetry with respect to 1$\pi$. So hereafter, we present the plot up to 1$\pi$. Next, we consider the onsite potential a cosine modulation $V=0.05\cos(2\pi bn)$, which suggests $V_A=-V_B=0.05$. The cosine modulation energy spectrum is plotted in Fig.\ref{fig5}(c) and splits into two bands, which opens a gap, and the edge states are split into two branches away from zero. As we continuously sweep the parameter $\phi$, the edge states mix with the bulk around 0.5$\pi$, and the different behaviour of the HHG spectrum is shown in Fig.\ref{fig5}(d).

\begin{figure}[t]
\centering\includegraphics[width=1.0\columnwidth]{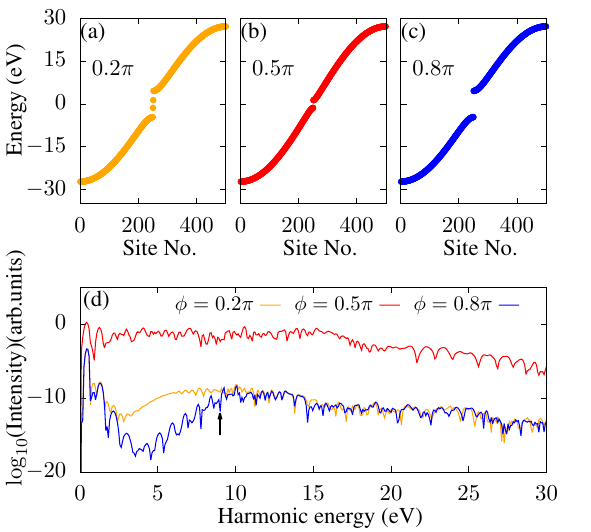}
\caption{The energy spectrum presented for the $V_A=-V_B$ case with three different $\phi$ values: (a) $\phi=0.2\pi$, (b) $\phi=0.5\pi$, and (c) $\phi=0.8\pi$, along with the harmonic emission spectra, are presented in (d). The black arrow represents the minimum bandgap between the VB and CB for $0.2\pi$ and $0.8\pi$.}
\label{fig6}
\end{figure}

\begin{figure}[b]
\centering\includegraphics[width=1.0\columnwidth]{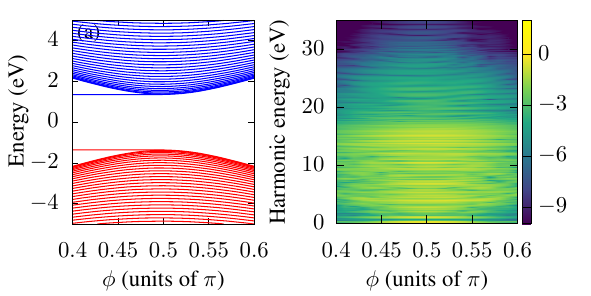}
\caption{The zoomed view of Fig.\ref{fig5}(c) in the range $\phi=0.4-0.6\pi$ is shown in (a), and Fig.\ref{fig5}(d) is presented in (b).}
\label{fig7}
\end{figure}

As discussed in Figs.\ref{fig5}(b,d), the harmonic emissions differ in the above cases. To further emphasise the impact of onsite potential in the HHG spectrum, we look into case $V_A=-V_B$, where the harmonic emission weakens for specific $\phi$ values as compared to the chiral preserving system, as shown in Fig.\ref{fig5}(d). To understand that, we have taken the three different $\phi$ values: $\phi=0.2\pi$, $\phi=0.5\pi$, and $\phi=0.8\pi$. The corresponding band structure and harmonic spectrum are in Fig.\ref{fig6}(a-d). We can see that harmonic intensity increases around 2.7 eV at $0.2\pi$, which corresponds to a transition between the valence band and the lower edge state. From here, interband transitions happen directly in the conduction band because they inhibit the transitions between the two edge states, which are localised at the opposite boundaries of the chains. This reflects the lower harmonic yield in the minimum bandgap region at 0.2$\pi$ as compared to chiral-preserving systems. The harmonic yield at 0.8$\pi$ is less than 0.2$\pi$, which confirms the absence of the edge state. The black arrow represents the minimum band gap for $0.2\pi$ and $0.8\pi$. The band structure for $0.5\pi$ is gapped, and the higher harmonic emission was observed because of the small bandgap compared to other $\phi$ values. In the $\phi$-dependent HHG spectrum (Fig.\ref{fig5}(d)), the intensity enhanced around $0.4\pi-0.6\pi$, and we have presented the energy and HHG spectrum for the same range as in Fig.\ref{fig7}(a). Throughout this range, there is a bandgap between the VB and CB, which dictates that the strength of the interband transitions increases, and the total harmonic spectrum dominates the interband harmonics. In contrast, the bandgap is reduced in the chiral preserving system around the phase transition region; the inter- and intraband oscillations are comparable. Hence, the interference term has a crucial impact on the total harmonic spectrum.

\begin{figure}[t]
\centering\includegraphics[width=1.0\columnwidth]{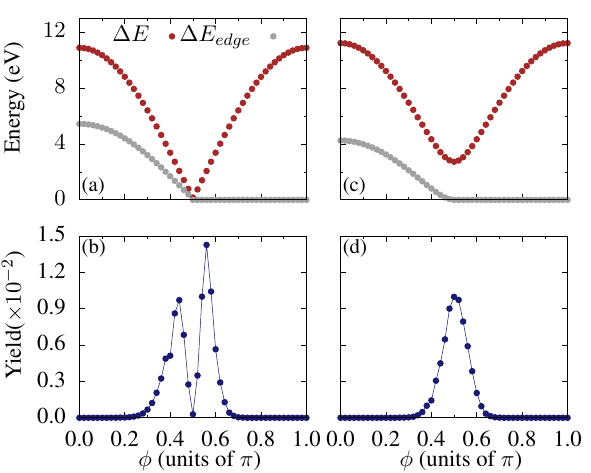}
\caption{$\Delta E$ is the bandgap between the VB and CB, and $\Delta E_{edge}$ is the bandgap defined as the VB and the edge state, which lies in the topological non-trivial phase as plotted with $\phi$ in (a). The harmonic yield in the 3–9 eV energy range for the off-diagonal hopping model with continuously scanning the modulating phase $\phi$ in (b). Similarly, $\Delta E$ (c) and the harmonic yield (d) are presented for adding onsite potential as $V_A=-V_B$.}
\label{fig8}
\end{figure}

\subsection{HHG Yield}  

\begin{figure}[t]
\centering\includegraphics[width=1.0\columnwidth]{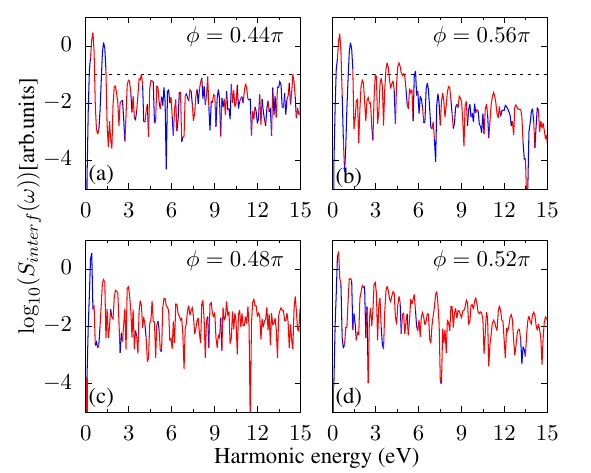}
\caption{Comparing the inteference term of $S_{\omega}$ between the topologically non-trivial phase at 0.44$\pi$ (a), 0.48$\pi$ (c), and the trivial phase at 0.56$\pi$ (b), 0.56$\pi$ (d) is plotted. The red and blue colours indicate destructive interference and constructive interference}
\label{fig9}
\end{figure}

Here, we will discuss the harmonic yield calculated as a function of $\phi$ in Fig.\ref{fig8}(b). The intercell hopping strength weakens, and the harmonic yield efficiency is enhanced by changing $\phi$ to 0.44$\pi$ after yield decreases until the topological phase transitions happen at 0.5$\pi$. As $\phi$ changes, the intracell hopping strength increases, and the bandgap opens. The $\Delta E$ is the energy difference between the VB and CB. In Fig. \ref{fig8}(a) shows that the $\Delta E$ has a mirror symmetry with respect to 0.5$\pi$ when plotted against $\phi$. The appreciable harmonic yield starts at 0.3$\pi$, which has the mid-energy state to enhance the transitions; after 0.6$\pi$, the harmonic yield rapidly decreases because the absence of mid-energy dictates the lower harmonic yield. To emphasise the fact that there is less harmonic yield after 0.44$\pi$ and an enhanced yield between $0.52\pi-0.58\pi$ as compared to these $0.48\pi-0.42\pi$. We have taken the two $\phi$ values, at 0.44$\pi$ and 0.56$\pi$, which represent the highest yield, and the corresponding interference term is presented in Fig.\ref{fig9}(a,b). The interference spectrum oscillates between constructive and destructive interference and is clearly visible in the colour plot for both $\phi$ values. The black dotted line in Fig.\ref{fig9}(a,b) shows that after 6 eV, interference strength at 0.56$\pi$ starts decreasing compared to 0.44$\pi$, which implies a higher yield at 0.56$\pi$. The interference terms for 0.48$\pi$ and 0.52$\pi$ are plotted in Fig.\ref{fig9}(c,d). The strong destructive interference seen in both the $\phi$ values leads to a reduction in the total spectra, resulting in a lower harmonic yield. Moreover, when there is a smaller bandgap, the interference effects play a dominant role in reducing the harmonic yield. The harmonic yield is a robust measure that allows us to distinguish emission spectra between the topologically trivial and non-trivial phases.

\begin{figure}[b]
\centering\includegraphics[width=1.0\columnwidth]{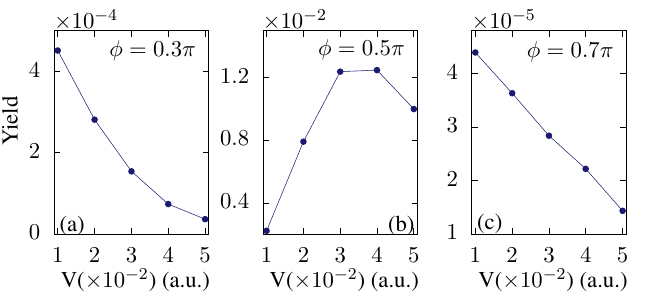}
\caption{The harmonic yield is calculated with the variation of the onsite strength "V" ranging from V=(0.01-0.05) a.u. for $V_A=-V_B$ and shown for three different $\phi$ values: (a) 0.3, (b) 0.5, and (c) 0.7. The scaling factors are specified in each plot.}
\label{fig10}
\end{figure}

\begin{figure}[b]
\centering\includegraphics[width=1\columnwidth]{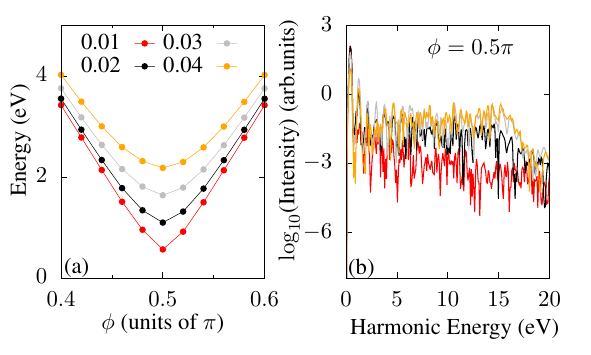}
\caption{The $\Delta E$ is the energy difference between bands (a) and HHG spectrum at 0.5$\pi$ (b) are presented with different onsite strengths for $V_A=-V_B$ case.} 
\label{fig11}
\end{figure}

Similarly, the harmonic yield for the broken chiral symmetry case is presented in Fig.\ref{fig8}(d). The harmonic yield for the $V_A=V_B$ case is not shown; it has just the shifted energy spectrum, and the yield is the same as in Fig.\ref{fig8}(b). The different onsite strength inside the sublattice is $V_A=-V_B$, which opens a gap in the energy spectrum. In Fig.\ref{fig8}(c), $\Delta E$ as a function of $\phi$ is plotted; the lifting of degenarate edge states implies it is not exactly located at half of the bandgap. The electron transition probability is less because the large bandgap between the first edge state and the VB leads to a lower efficiency before 0.38$\pi$, as presented in Fig.\ref{fig8}(d), and the efficiency is enhanced in the range $0.4\pi-0.6\pi$ that implicitly relates to the minimum bandgap. The position of the edge state plays a crucial role in strengthening  the harmonic emission. Breaking the chiral symmetry in this case reduces the window of harmonic emissions.

So far, we have learned that the efficiency of the harmonic yield is a good measure to identify the topological phases. If the harmonic yield increases or decreases, that relates to the minimum or maximum bandgap and also to the position of the edge states. In order to elucidate the lower harmonic yield obtained in the $V_A=-V_B$ case in Fig.\ref{fig8}(d), we have presented the harmonic yield with the variation of the onsite strength ranging from 0.01 to 0.05 a.u. It can be seen from Fig.\ref{fig10}(a-c) that as the strength of the onsite increases, the efficiency of the harmonic yield decreases, which correlates to the increase in the bandgap between the VB and CB for the $\phi$ values of 0.3$\pi$ and 0.7$\pi$ and conversely increasing in the harmonic yield with respect to onsite strength at 0.5$\pi$. In order to verify the harmonic yield enhancement for $\phi = 0.5\pi$ with different onsite potential of the form $V_A=-V_B$, we have presented the variation of the bandgap energy with $\phi$ for different onsite potential strength [Fig. \ref{fig11}(a)]. It clearly shows that the bandgap is varying significantly against onsite strengths. The respective HHG spectra for the case of  $\phi = 0.5\pi$ for different onsite potentials is also presented [Fig. \ref{fig11}(b)]. We have already seen that the minimum bandgap leads to destructive interference and that reduces the total HHG spectra. This explains the higher yield when the onsite strength increases at 0.5$\pi$. The above analysis shows that at 0.5$\pi$, there is an interesting feature in the variation of onsite strength. Hence, the harmonic yield detects all the aspects of the AAH model.

\section{Summary}
\label{sec4}

In summary, we have investigated the high-harmonic generation in the different topological phases of  off-diagonal hopping in the AAH model with a commensurate modulation, which gives rise to a zero-energy edge state by keeping the chiral symmetry intact. The harmonic spectra show the distinct features for topological nontrivial, trivial, and transition phases with a periodically modulated control parameter $\phi$. We calculated the intra- and inter-band harmonics around the phase transition point to identify the lowest harmonic yield. By comparing the total harmonic spectra, we observe that the interference term dominates and negates the contribution from intra- and inter-band harmonics. In addition, the impact of chiral symmetry breaking in the HHG spectrum is studied by adding the onsite potential in two different ways: $V_A=V_B$ and $V_A=-V_B$, and the position of the edge states found to play a crucial role in harmonic enhancement. The harmonic yield is calculated as a function of the control parameter $\phi$. Around the phase transitions, the yield is found to be minimum, implying the interference term has a dominant role. The yield is a very sensitive probe that distinguishes between the topologically nontrivial and the topologically trivial phases. Our work presents the HHG yield as a robust measure to detect topological phase transitions. Furthermore, the onsite strength can control the harmonic emission of the systems with broken chiral symmetry. Explicit dependence of HHG processes on higher dimensions and the decoherence we reserve for the future. 
 
%\bibliography{Ref} % Use sample.bib as the bibliography file


\end{document}